l-aa.dem
\documentstyle{l-aa}

\begin{document}

   \thesaurus{ (08.14.1; 
               08.16.6;  
               13.07.1)} 

   \title{Gamma-ray burst afterglows and evolution of postburst fireballs
           with energy injection from strongly magnetic millisecond pulsars}

      \author{Z. G. Dai$^1$ and T. Lu$^{2,1,3}$}

   \offprints{Z. G. Dai}

   \institute{$^1$Department of Astronomy, Nanjing University, 
                   Nanjing 210093, P.R. China \\
              $^2$CCAST (World Laboratory), P.O. Box 8730, 
                   Beijing 100080, P.R. China \\
              $^3$LCRHEA, IHEP, CAS, Beijing, P.R. China}

   \date{Received ???; accepted ???}

   \maketitle\markboth{Z.G. Dai \& T. Lu: Gamma-ray burst afterglows 
                       and evolution of fireballs with pulsars}{}

   \begin{abstract}

Millisecond pulsars with strong magnetic fields may be formed through
several processes, e.g. accretion-induced collapse of magnetized
white dwarfs, merger of two neutron stars, and accretion-induced phase
transitions of neutron stars. During the birth of such a pulsar, an initial
fireball available for a gamma-ray burst (GRB) may occur. We here study
evolution of a postburst relativistic fireball with energy injection from
the pulsar through magnetic dipole radiation, and find that the magnitude of
the optical afterglow from this fireball first decreases with time,
subsequently flattens, and finally declines again. This may provide
a natural explanation for the behavior of the lightcurve of
the afterglow of GRB970228 if this burst resulted from
the birth of a strongly magnetic millisecond pulsar.


   \end{abstract}

%

\section{Introduction}

   The observational results from the BATSE detector on {\em CGRO}
(Fishman \& Meegan 1995) strongly suggested that the sources of gamma-ray 
bursts (GRBs) are at cosmological distances or in the galactic halo,
leading to a great debate on the origin of GRBs. The recent discovery of
fading sources at X-ray (Costa, et al. 1997a, b; Yoshida, et al. 1997) and
optical wavebands (Groot, et al. 1997; van Paradijs, et al. 1997;
Sahu, et al. 1997a, b, c), and the detection of the redshift of the
counterpart 
to GRB970508 (Metzger, et al. 1997), clearly demonstrate that GRBs are 
at cosmological distances.

Most of the models proposed to explain the occurrence of cosmological GRBs
are related with compact objects from which a large amount of energy
available for an extremely relativistic fireball is extracted through neutrino
emission and/or dissipation of electromagnetic energy. One such possibility
is newborn millisecond pulsars with strong magnetic fields, which may be
formed by several models, e.g. (1) accretion-induced
collapse of magnetized white dwarfs, (2) merger of two neutron stars, and
(3) accretion-induced phase transitions of neutron stars. In the first
model, an accreting white dwarf, when its mass increases up to
the Chandrasekhar limit, may collapse to a rapidly rotating neutron star,
and the magnetic field of the neutron star may become very strong
due to magnetic-flux conservation (Usov 1992) or efficient dynamo action
(Duncan \& Thompson 1992). During the birth of the neutron star,
an initial fireball may occur through neutrino processes (Dar, et al. 1992)
and/or electromagnetic processes (Usov 1992; Yi \& Blackman 1997).

Second, it is usually thought that a neutron-star binary
will eventually merge into a black hole due to gravitational radiation.
However, since the equations of state at high densities are likely stiff for 
several observational and theoretical facts summarized by Cheng \& Dai
(1997), Dai \& Cheng (1997), and Cheng \& Dai (1998),
neutron-star binaries like the Hulse-Taylor system might merge into
massive neutron stars which have both millisecond periods and very strong
magnetic fields generated by some physical processes, e.g. dynamo action
powered by tidal heating (Vietri 1996). During the coalescence and merging of
two neutron stars, an initial fireball may be produced through
neutrino-antineutrino annihilation (Eichler, et al. 1989; Mathews, et al.
1997)
and/or efficient energy transfer from the orbital energy
(Vietri 1996). Third, it has been proposed by Cheng \& Dai (1996) that
some neutron stars in low-mass X-ray binaries accrete sufficient mass
to undergo phase transitions to become strange stars, and after the 
phase transitions fireballs of GRBs naturally occur from the quark 
surfaces of the hot strange stars (Cheng \& Dai 1996; Usov 1998) because of 
very-low-mass crusts. The resulting strange stars may have short 
periods due to accretion-induced spin-up prior to the phase transitions,
and they could have strong magnetic fields (Cheng \& Dai 1997).
Therefore, it can be seen that two common products of these three
energy-source models are newborn millisecond pulsars with strong magnetic
fields and initial fireballs available for GRBs. Furthermore, Klu\'zniak 
\& Ruderman (1997) proposed that strongly magnetized neutron stars 
differentially rotating at millisecond periods may be the central engine
of gamma-ray bursts.

It is natural to expect that if a GRB results from the birth of
a strongly magnetic millisecond pulsar, then after the main GRB
the pulsar continuously supplies energy to a relativistic fireball through
magnetic dipole radiation. In this Letter we study evolution of the radiation 
from such a postburst fireball. 

%

\section{The model}

We assume, as a pulsar is born by the models summarized in the introduction,
an amount of energy $E$ comparable to that observed in gamma rays,
$E\sim 10^{51}$ ergs, is released over a time interval less than 100 seconds.
This initial fireball will expand and accelerate to relativistic velocity
because of the huge optical depth in the source (Paczy\'nski 1986;
Goodman 1986; Shemi \& Piran 1990). Perhaps, internal shocks
are formed due to collisions between the shells with different Lorentz
factors in this period (Rees \& M\'esz\'aros 1994; Paczy\'nski \& Xu 1994). 
During the acceleration, the initial radiation energy will be
converted to bulk kinetic energy of the outflow. Subsequently,
the expansion of the fireball starts to be significantly influenced by the
swept-up interstellar medium (ISM) and two shocks will be formed: a forward
blastwave and a reverse shock (Rees \& M\'esz\'aros 1992; Katz 1994).
A GRB will be produced once the kinetic energy is dissipated and radiated
as gamma rays through synchrotron or possibly inverse-Compton emission from
the accelerated electrons in the shocks (M\'esz\'aros, et al.
1994; Sari, et al. 1996). The postburst fireball
will be decelerated continuously, and an X-ray, optical and/or radio 
afterglow will be formed, as predicted early by many authors (Paczy\'nski 
\& Rhoads 1993; Katz 1994; Vietri 1997b; M\'esz\'aros \& Rees 1997).

At the center of the fireball, the pulsar loses its
rotational energy through magnetic dipole radiation, whose power is
given by 
\begin{eqnarray}
L & = & \frac{2}{3c^3}\left(\frac{2\pi}{P}\right)^4 R^6 B_s^2\sin^2\theta
\nonumber  \\
& = & 4\times 10^{43}\,{\rm ergs}\,{\rm s}^{-1}B_{\bot,12}^2P_{\rm ms}^{-4}
R_6^6\,,
\end{eqnarray}
where $B_{\bot,12}=B_s\sin\theta/10^{12}\,{\rm G}$, $B_s$ is the surface
dipole field strength, $\theta$ is the angle between the magnetic dipole
moment and rotation axis, $P_{\rm ms}$ is the period in units of 1 ms, and 
$R_6$ is the stellar radius in units of $10^6\,$cm. This power varies 
with time as 
\begin{equation}
L(t) \propto (1+t/T)^{-2}\,, 
\end{equation}
where $t$ is one measure of time in the burster's rest frame,
and $T$ is the initial spin-down timescale defined by 
\begin{equation}
T \equiv \left(\frac{P}{2\dot{P}}\right)_0
= 5\times 10^8\,{\rm s}\,B_{\bot,12}^{-2}P_{\rm ms}^2I_{45}R_6^{-6}\,,
\end{equation}
where $\dot{P}$ is the rate of period increase due to magnetic dipole 
radiation and $I_{45}$ is the stellar moment of inertia in units of
$10^{45}\,{\rm g}\,{\rm cm}^2$. (Note that the stellar
spin-down timescale in the burster's rest frame is equal to that in the
observer's frame.) For $t< T$, $L$ can be thought as a constant;
but for $t\gg T$, $L$ decays as $\propto t^{-2}$.

The power of the pulsar is radiated
away mainly through electromagnetic waves with frequency of $\omega=2\pi/P$.
Once the electromagnetic waves propagate in the shocked ISM, they will
be absorbed if $\omega_p > \omega$, where $\omega_p$ is the plasma
frequency of the shocked ISM. Since the electron number density of
the shocked ISM in the burster's rest frame is $n_e=4\gamma^2n$
(Blandford \& McKee 1976), where $\gamma$ is the Lorentz factor of
the shocked ISM and $n$ is the electron number density of the unshocked ISM,
for $\omega\sim 10^4\,{\rm s}^{-1}$ the inequality $\omega_p>\omega$
is always valid if $n>0.01\gamma^{-2}\,{\rm cm}^{-3}$. This shows that
the electromagnetic waves cannot propagate through the shocked ISM and
by this means energy can be continuously pumped from the pulsar
into the shocked ISM. This idea is similar to that of Pacini (1967), 
who first proposed that a pulsar can continuously supply energy 
to its surrounding supernova remnant through magnetic dipole radiation.

In the following we assume that the expansion of the postburst blastwave in
uniform ISM is relativistic and adiabatic. At a time $t$, the shocked
ISM energy is given by $E_{\rm sh}=4\pi r^2(r/4\gamma^2)\gamma^2e'$
(Waxman 1997a), where $r\approx ct$ is the blastwave radius and 
$e'=4\gamma^2nm_pc^2$ is the shocked ISM 
energy density in the comoving frame (Blandford \& McKee 
1976). This energy should be equal to the sum of $E/2$ and the energy which 
the fireball has obtained from the pulsar based on energy conservation:
\begin{equation}
4\pi nm_pc^2\gamma^2r^3=\frac{E}{2}+\int_0^t (1-\beta)L(t-r/c)dt\,,
\end{equation}
where $\beta=(1-1/\gamma^2)^{1/2}$ and the factor $1/2$ arises from the
fact that about half of the total energy of the initial fireball has been
radiated away during the GRB phase (Sari \& Piran 1995).
The term $(1-\beta)$ accounts for the fact that the fireball moves away
from the incoming photons, and the power $L(t-r/c)$ shows the fact
that the radiation absorbed by the fireball at time $t$ was emitted at
the retarded time $t-r/c$. According to Eq. (2), this power should
decrease as $L(t-r/c)\propto [1+(t-r/c)/T]^{-2}=(1+t_\oplus/T)^{-2}$,
where $t_\oplus$ is the observed time. Because the Lorentz factor of
the fireball at the initial evolution stage decays as $\gamma\propto
t^{-3/2}$, the timescale in which the fireball has obtained energy $\sim E/2$
from the pulsar can be estimated by $4\gamma^2E/L$, which is measured in
the burster's rest frame. The corresponding observer-frame timescale ($\tau$)
is equal to this timescale divided by $2\gamma^2$ (Waxman 1997b), viz., 
$\tau=2E/L$. We require $\tau \sim 6$\,days and $T\gg \tau$ for the
optical afterglow of GRB970228.

We now analyze evolution of the afterglow from a postburst fireball with
energy injection from the pulsar through magnetic dipole radiation.
First, at the initial stage of the afterglow, viz., $t_\oplus\ll \tau$,
the first term on the right hand of Eq. (4) is much larger than the
second term. In this case, one expects that the fireball is not
significantly influenced by the stellar radiation, and thus its expansion 
evolves based on $\gamma=324E_{51}^{1/8}n_1^{-1/8}t_\oplus^{-3/8}\,$,
where $E_{51}=E/10^{51}\,{\rm ergs}$, $n_1=n/1\,{\rm cm}^{-3}$,
and $t_\oplus$ is in units of 1 second.
According to Dai \& Lu (1997), therefore, the synchrotron flux (X-ray)
integrated between 2 and 10 keV decays as $F_X\propto t_\oplus^{-\alpha_X}$,
where $\alpha_X=3/2-3(3-p)/16$, and the synchrotron flux density
at some optical band
declines as $S_{\rm opt}\propto t_\oplus^{-\alpha_{\rm opt}}$, where
$\alpha_{\rm opt}=3(p-1)/4$. Here $p$ is the index of the power-law
distribution of the accelerated electrons in the shocked ISM.
The observations of the afterglow of GRB970228 have given
$\alpha_X=1.4\pm 0.2$ (Costa, et al. 1997a, b;
Yoshida, et al. 1997) and $\alpha_{\rm opt}=2.1^{+0.3}_{-0.5}$ in 
R band (Galama, et al. 1997). In order to fit these values, we require
$p\sim 3$.

Second, for $T>t_\oplus\gg \tau$,
according to Eq. (4), the energy which the fireball has
obtained from the pulsar is much larger than $E/2$, and the expansion of
the fireball is significantly affected by the pulsar's radiation. In the 
case with steady energy supply, the Lorentz factor of the fireball decays 
as $\gamma\propto r^{-1/2}\propto t_\oplus^{-1/4}$ (Blandford \& McKee 1976). 
This scaling law can also be found by neglecting the term $E/2$ in Eq. 
(4). Here we have assumed that the power radiated from the pulsar doesn't 
vary with time. The comoving-frame equipartition magnetic field decays as
$B'\propto \gamma\propto t_\oplus^{-1/4}$, and the synchrotron
break frequency drops in time as $\nu_m\propto \gamma^3B'\propto
t_\oplus^{-1}$. At the same time, since the comoving electron
number density is $n'_e\propto \gamma\propto t_\oplus^{-1/4}$ and
the comoving width of the emission region $\Delta\! r' \sim r/\gamma
\propto t_\oplus^{3/4}$, according to M\'esz\'aros \& Rees (1997a) and
Wijers et al. (1997), the comoving intensity $I'_\nu\propto
n'_eB'\Delta\! r'\propto t_\oplus^{1/4}$. So the observed peak flux density
increases in time based on $S_{\nu_m}\propto t_\oplus^2 \gamma^5I'_{\nu_m}
\propto t_\oplus$. Thus, for $\nu\gg\nu_m$, the observed optical flux 
density as a function of observed time is
\begin{equation}
S_\nu=S_{\nu_m}(\nu/\nu_m)^{-(p-1)/2}\propto t_\oplus^{(3-p)/2}\,.
\end{equation}
Therefore, the requirement of $p\sim 3$ inferred from the early-time
behavior of the afterglow leads to the result that the observed R-band flux
at the subsequent stage may keep relatively constant.

Third, for $t_\oplus\gg T$, the power of the pulsar due to magnetic 
dipole radiation rapidly decreases as $L\propto t_\oplus^{-2}$, and
thus the fireball is hardly influenced by the stellar radiation and the
optical flux of the afterglow will significantly decline with time again.
  
%

\section{Constraints on stellar parameters}

The analysis of the lightcurve in the last section
can be directly applied to discussion on the afterglow of GRB970228.
As pointed out by Galama et al. (1997),
the initial decrease of the optical afterglow can be easily
explained by the popular fireball model (M\'esz\'aros \& Rees 1997;
Wijers, et al. 1997; Waxman 1997a; Vietri 1997a; Tavani 1997; Reichart 1997; 
Dai \& Lu 1997), but the subsequent flattening is not consistent
with this simple model. Further considering the result observed by HST on
4 September (Fruchter, et al. 1997), therefore, we can see
that the magnitude of the R-band brightness of the afterglow of GRB970228
first rapidly decreased with time, subsequently flattened, and finally
declined again. The lightcurve analyzed in the last section is
well consistent with these observational results.

If the afterglow of GRB970228 was radiated from a 
relativistic fireball with energy injection from a pulsar through
magnetic dipole radiation, we now give constraints on some parameters
of this pulsar. Inserting Eq. (1) into $\tau=2E/L$,
we get the timescale at which the fireball has obtained energy $\sim E/2$:
\begin{equation}
\tau \approx 5\times 10^7\,{\rm s}\,E_{51}B_{\bot,12}^{-2}P_{\rm
ms}^4R_6^{-6}\,.
\end{equation}
This equation can be further changed into the following form:
\begin{equation}
B_{\bot,12}\approx 10E_{51}^{1/2}(\tau/6\,
{\rm d})^{-1/2}P_{\rm ms}^2 R_6^{-3}\,.
\end{equation}

The R-band magnitude of the optical afterglow
during one month following 6 March after GRB970228 didn't vary with time
(Galama, et al. 1997), which means that during this period the stellar
radiation significantly affected the evolution of the fireball.
This in fact requires one condition: $T\ge 36$ days. On the other hand,
the V-band magnitude of the afterglow on 4 September dropped to
$V=28.0\pm 0.25$ which corresponds to the R-band magnitude $R\approx 27.0$
(Fruchter, et al. 1997), showing that for $t_\oplus \gg 1$ month
the stellar radiation played no role in the expansion of the fireball.
This requires another condition: $T\ll 188$ days. Furthermore, after
comparing the observed data with the decay law of the optical flux at
the early stage or at the late stage ($S_{\rm opt}\propto 
t_\oplus^{-3/2}$), we can infer that the flattening phase of the optical 
flux must have ended around 60 days. This means $T\le 60\,$ days. 
From these conditions and Eqs. (3) and (7), we obtain
\begin{equation}
1.0\le P_{\rm ms}E_{51}^{1/2}(\tau/6\,{\rm d})^{-1/2}I_{45}^{-1/2}\le 1.2\,.
\end{equation}
The equations of state at high densities are likely stiff
and thus the moments of inertia of rapidly 
rotating neutron stars with mass $\ge 1.4M_\odot$ are $I_{45}\sim 2$
(Datta 1988; Weber \& Glendenning 1993). Further adopting $R_6\sim 1$,
$E_{51}\sim 1$ and $\tau\sim 6$\,days, we can find $1.4\le P_{\rm ms}\le 1.7$
and $20\le B_{\bot,12}\le 30$. Therefore, we suggest that the central engine
of GRB970228 could be a strongly magnetic millisecond pulsar.

%

\section{Discussion}

We have studied evolution of a postburst fireball with energy injection
from a millisecond pulsar with a strong magnetic field through
magnetic dipole radiation if the initial fireball of a GRB has been produced
during the birth of the pulsar. In the case of $\tau < T$ or
$P_{\rm ms}<3.2\,{\rm ms}I_{45}^{1/2}E_{51}^{-1/2}$, according to
Eqs. (3) and (6), we found that the magnitude of
the optical afterglow from this fireball first decreases with time,
subsequently flattens, and finally declines again. We have
assumed a newborn millisecond pulsar with a strong magnetic
field to be an origin of GRB970228, but have not invoked any specific
model. There may be several mechanisms relating the birth
of strongly magnetic millisecond pulsars with GRBs, e.g. accretion-induced
collapse of magnetized white dwarfs, merger of neutron-star binaries,
and accretion-induced phase transitions of neutron stars.
One expects that one of them could be the mechanism of the birth of
a strongly magnetic millisecond pulsar if this pulsar is the central engine
of GRB970228. In the case of $\tau > T$ or $P_{\rm ms}>3.2\,{\rm ms}I_{45}
^{1/2}E_{51}^{-1/2}$, the millisecond pulsar as the central engine cannot
supply energy to the postburst fireball effectively enough for the
flattening phase to occur. It is interesting to note that this feature
doesn't depend on magnetic field strength, which influences $\tau$ and $T$
in the same way.

We further assume that {\em some} cosmological GRBs result from the birth of
strongly magnetic millisecond pulsars. If the effective 
dipole magnetic-field strength 
of a newborn pulsar is extremely high (e.g., $B_\bot =B_s\sin\theta \sim 
10^{15}\,$G), the stellar rotational energy is dissipated due to magnetic 
dipole radiation in observed time $T<10^2\,$s. In other words, the 
fireball of a GRB can obtain a large amount of energy from the pulsar only
in such a short timescale. For $t_\oplus\gg T$, thus, the expansion 
of the postburst fireball is hardly affected by the magnetic dipole radiation 
of the pulsar. In the superstrong-magnetic-field case, the flattening
behavior might not be observable in the optical counterparts of GRBs.

\begin{acknowledgements}
    We would like to thank the referee for very valuable
suggestions, and Drs. J. M. Wang and D. M. Wei for many discussions.
This work was supported by the National Natural Science Foundation of China.  
\end{acknowledgements}

\end{document}